\documentclass{pasj00}
\def\gsim{\ \raise 3pt \hbox{$>$} \kern -8.5pt \raise -2pt \hbox{$\sim$}\ }

\begin{document}
\SetRunningHead{Fleishman}{Particle Acceleration in Flares}
\Received{2013/03/24}
\Accepted{2013/03/31}

\title{Microwave View on Particle Acceleration in Flares}

\author{Gregory D. \textsc{Fleishman}}%
\affil{Physics Dept., New Jersey Institute of Technology, Newark, MJ
07102 USA} 

\affil{Central Astronomical Observatory at Pulkovo of RAS, Saint-Petersburg 196140, Russia} 

%

\maketitle

\begin{abstract}
Although the solar flare phenomenon is widely accepted to be a consequence of release of excessive magnetic energy stored in the coronal currents (stated another way---in nonpotential magnetic fields), many essential details of this energy release remain poorly understood. Initially, the released flare energy is somehow divided between thermal and nonthermal components through plasma heating and particle acceleration, respectively, although this proportion can then change in the course of the flare due, e.g., to fast particle Coulomb losses leading to additional plasma heating and/or chromospheric evaporation. So far, the thermal-to-nonthermal partition was found to vary greatly from one flare to another resulting in a broad variety of cases from 'heating without acceleration' (Battaglia et al. 2009) to 'acceleration without heating' (Fleishman et al. 2011). Recent analysis of microwave data of these differing cases suggests that a similar acceleration mechanism, forming a power-law nonthermal tail up to a few MeV or even higher, operates in all the cases. However, the level of this nonthermal spectrum compared to the original thermal distribution differs significantly from one case to another, implying a highly different thermal-to-nonthermal energy partition in various cases. This further requires a specific mechanism capable of extracting the charged particles from the thermal pool and supplying them to a bulk acceleration process to operate in flares \textit{in addition} to the bulk acceleration process itself, which, in contrast, efficiently accelerates the seed particles, while cannot accelerate the thermal particles. Within this 'microwave' view on the flare energy partition and particle acceleration I   present a few contrasting examples of acceleration regions detected with microwave data and compare them with  the most popular acceleration mechanisms---in DC fields, in collapsing traps, and stochastic acceleration by a turbulence spectrum---to identify the key elements needed to conform with observations. In particular, I point out that the turbulence needed to drive the particle acceleration is generated in nonpotential magnetic structures, which results in nonzero helicity of the turbulence. This helicity, in its turn, produces a nonzero mean DC electric field on top of stochastic turbulent fields driving the main stochastic acceleration; thus, acceleration by helical turbulence combines properties of the standard stochastic acceleration with some features of acceleration in DC electric fields, exactly what is demanded by observation. 
\end{abstract}

\section{Introduction}

Microwave continuum radio bursts are believed to be primarily produced by an electron population magnetically trapped near the top of a flaring magnetic loop rather than the accelerated electron component directly \citep{Lee_Gary_Zirin_1994, Melnikov_1994, Kundu_etal_2001, melnikov_etal_2002}. Using the radio emission produced by this trapped electrons for diagnostics of the particle acceleration is difficult  because this task requires to disentangle effects of acceleration and transport, which is expected to be model dependent. Although some exciting results have been obtained in this way (see, e.g., \cite{Reznikova_etal_2009}), it is yet unclear how unique they are. The most recent developments in this area are discussed in Melnikov's article in this proceeding volume, so I am not going to discuss the particle transport outside the acceleration region in any detail.

To outline the framework of the further discussion we emphasize that the GS continuum radio emission can be produced by any of (i) a magnetically trapped component or (ii) a  precipitating component, or (iii) the primary component within the acceleration region, rather than exclusively by the magnetically trapped component.
What is highly important for diagnostics, these three populations of fast electrons produce radio emission with distinctly different characteristics \citep{Fl_etal_2011}. Indeed, in the case of magnetic trapping the electrons are accumulated at the looptop (Melnikov et al. 2002), and the radio light curves must be delayed by roughly the trapping time relative to accelerator/X-ray light curves.  In the case of radio emission from the acceleration region, even though the residence time that fast electrons spend in the acceleration region can be relatively long, the radio and X-ray light curves are proportional to each other simply because the flux of the X-ray producing electrons is equivalent to the electron loss rate from the acceleration region.

Therefore, what is needed to study the acceleration region in the microwave domain is to cleanly separate its contribution from the two other mentioned competing contributions---from the magnetically trapped and precipitating components. Note that having imaging spectroscopy observations this separation will be routinely possible for many events. Meanwhile, however, we are limited to some favorable cases when either no magnetic trapping takes place or the acceleration region contribution dominates temporarily or/and spectrally over the competing contributions. Below we review a few such favorable cases and discuss the obtained acceleration region properties vs available mechanisms of particle acceleration.

Main acceleration mechanisms that can play a role in solar flares include both regular and stochastic processes.
Regular energy gain can take place in a DC electric field or in a contracting source (e.g., collapsing magnetic trap), while the stochastic acceleration can be driven by turbulence either resonantly (the case of short-wave turbulence) or nonresonantly (large-scale turbulent pulsations). There can be processes combining some regular and stochastic features---e.g., diffusive shock acceleration, while acceleration by an ensemble of shock waves represents a special example of stochastic acceleration (for general overview of the acceleration processes see, e.g., \cite{Fl_Topt_2013_CED}). Importantly, predicted observational manifestations of all these acceleration mechanisms are different from each other \citep{Li_Fl_2009, Park_Fl_2010}, which implies that they can be distinguished observationally.

\section{Cold, dense flare: acceleration with mild heating}

Bastian et al. (2007) analyzed Yokhoh and GOES X-ray data together with Nobeyama and
OVSA radio observations to describe a new class of 'cold flares'---flares that occur in such
dense coronal magnetic loops that the flare energy deposition is insufficient to substantially
heat the coronal plasma, even though most of the flare energy is deposited in the coronal
rather than the chromospheric part of the dense loop.

A 'textbook' example of this class of events was the solar flare occurred on 2001 October 24 in
NOAA active region 9672 at a heliocentric position of S18W13 from
approximately 23:10-23:14 UT. A strong radio burst
was well-observed in total intensity by OVSA, NoRP, and NoRH
allowing a detailed analysis of the event. A puzzling feature of
this flare is that no soft X-ray emission was noted by NOAA/SEC
from the event, although weak extreme-ultraviolet (EUV), soft
X-ray (SXR), and hard X-ray (HXR) emissions were detected from the
flare  by the {\sl Transition Region and Coronal Explorer} (TRACE;
Handy et al. 1999), the {\sl Yohkoh} Soft X-ray Telescope (SXT),
and Hard X-ray Telescope (HXT), respectively (Kosugi et al. 1992),
as was a weak SXR enhancement by the GOES 10 satellite.

\begin{figure} [htbp]
\qquad \includegraphics[width=0.9\columnwidth,clip]{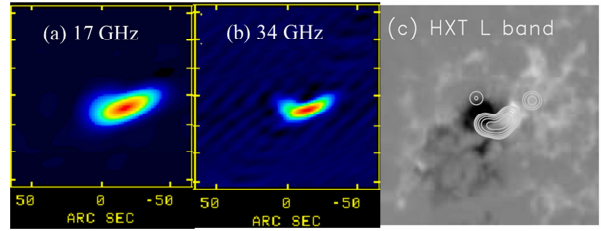}
\caption{
Flare images and the Kitt
Peak magnetogram. a,b) The images of the 17 and 34 GHz total intensity at the
time of 17 GHz flux maximum. c) HXT L band map on top of the Kitt
Peak magnetogram.}\label{norh_images}
\end{figure}

Thus, the microwave data is the main source of information for this
event. Fig.~\ref{norh_images}a,b shows the 17 and 34 GHz maps in
total intensity (Stokes I) near the time of the emission maximum. The source morphology evolves very
little at radio wavelengths as a function of time. The peak
brightness temperature of the 17 GHz source is $4.6\times 10^7$ K
whereas that of the 34 GHz source is $1.8\times 10^7$ K.
Fig.~\ref{norh_images}c shows the HXR source in HXT/Yohkoh L band superposed on the magnetogram. The
source in the SXR (not shown for brevity), HXR, and radio bands are coincident and quite
similar in morphology, therefore, all imaging data are consistent
with the illumination of a simple dense magnetic loop. The
loop is not visible in the NoRH 17 and 34 GHz maps as a discrete
feature in the active region prior to the flare; the estimate of the
17 GHz brightness temperature prior to the flare is
$\lesssim\!10^5$ K.


\begin{figure} [!h]
\includegraphics[width=0.95\columnwidth]{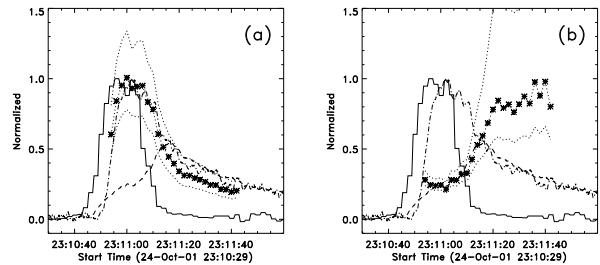}
\caption{The variation of $n_{rl}$ and $T$ with time derived from the fits
in comparison to the HXR (solid line) and radio
emission at 9.4 (dashed line) and 17 GHz (dash-dotted line). a) The asterisks show the variation of
fitted value of $n_{rl}$ (divided by $10^7$ cm$^{-3}$) for an
assumed magnetic field  of 165 G in the source. The dotted lines
above and below this line are the corresponding values for the
magnetic field strengths of 150 and 180 G, respectively. b) The
corresponding plot for the temperature $T$ variations (normalized by 6~MK) for the same
assumption. \  }\label{Cool_fit} 
\end{figure}

The spectrum of the radio emission with a sharp cutoff in the emission below
$\approx10$ GHz, while a hard slope above $\approx20$ GHz,
is clearly
nonthermal, yet the brightness temperature of the 17 and 34 GHz
sources is unexceptional, characteristic of optically thin
gyrosynchrotron emission. It is reasonable to conclude that the low frequency
cutoff is the result of the Razin effect, which strongly
suppresses the gyrosynchrotron emission in the presence of an
ambient plasma below a cutoff frequency $\nu_R\approx
20n_e/B_\perp$, where $n_e$ is the thermal electron density and
$B_\perp$ is the perpendicular component of the magnetic field
vector in the source relative to the line of sight. For $\nu_R\sim
10-15$ GHz, a magnetic field strength of $\sim\!150-200$ G at an
angle $\theta=60^\circ$ to the line of site implies the density is
$n_e\sim 10^{11}$ cm$^{-3}$. Higher magnetic fields imply higher
densities.

Estimate of the free-free optical depth  at radio wavelengths
shows
that if Razin suppression is important, free-free absorption is,
too. \citet{Bastian_etal_2007} concluded that a combination of \textit{Razin suppression}
and \emph{free-free absorption} plays a role in determining the
shape of the radio spectrum.

\citet{Bastian_etal_2007} applied a nonlinear model-fitting code that
adjusts model parameters to minimize the $\chi^2$ statistic using
the downhill simplex method \citep{Press_etal_1986}.
They fit the model to 25 composite radio spectra observed by OVSA
(5-14.8 GHz) and the NoRP (9.4, 17, 35, and 80 GHz),
as indicated in Fig.~\ref{Cool_fit}.

A uniform source model with an area $A=2\times
10^{18}$ cm ($12"\times 30"$) and a depth $L=9\times 10^8$ cm
(12"), consistent with the X-ray and radio imaging was used.
The source
volume is assumed to contain thermal background plasma with a
density $n_{th}$ and a temperature $T$. The source is assumed to
be permeated by a coronal magnetic field $B$ with an angle
$\theta$ relative to the line of sight. A power-law distribution
of energetic electrons $N(E)dE=KE^{-\delta}dE$ is assumed, with a
normalization energy $E_o$ and a high-edge cutoff energy $E_c$.
The total number density of energetic electrons between $E_o$ and
$E_c$ is $n_{rl}$.

Fig.~\ref{Cool_fit}   shows a comparison between the
fitted values for $n_{rl}$ and $T$ and the radio and HXR emission
as a function of time for a magnetic field of
165~G.  Two points can be made about the apparent
trends: 1) the total number of energetic electrons tracks the
variation of the radio emission. The maximum of the $n_{rl}$ is
coincident (to within 2 s) to the 35 GHz flux maximum; 2) the
temperature of the ambient plasma increases with time. With
$B=165$ G, $T$ increases from $\sim 1-2\times 10^6$ K to only $\sim 4-6
\times 10^6$ K. This manifests a discovery of a new class of flares, which occur in
relatively compact cool and dense loop, giving rise to efficient
acceleration of relativistic electrons and strong microwave
emission, although to only a moderate plasma heating and very week
X-ray emission.

Nevertheless, the modest coronal
plasma heating can be sensitively measured via a corresponding decrease of the free-free
radio opacity of the flaring plasma. This permits a precise
calorimetry of the flare energy deposited into nonthermal electrons and dissipated in the
corona to heat the ambient plasma. From this analysis we found that the ratio of total
energy of the nonthermal electrons in this flare was comparable to ($\sim 30\%$ of) the magnetic
energy of the flaring loop. The conclusion that the accelerated electron energy is about a
few tens percent of the flare energy is consistent with recent RHESSI results (Brown et al.
2007) on the nonthermal flare energy budget.

\citet{Bastian_etal_2007} analyzed the decay phase of the radio light curves and found that the exponential decay constants did not depend on the frequency, which means that the particle escape time did not depend on the particle energy. Such an energy-independent particle transport implies that the transport was mediated by turbulence rather than the Coulomb collisions; it is the turbulence that could drive stochastic electron acceleration at the impulsive phase of the flare. Note, that an interesting property of this transport/acceleration regime is the observationally-proven \textit{independence} of the \textit{electron escape time on energy}.


\section{Cold, tenuous flare: acceleration without heating}

\begin{figure}[!h]
\centering
\vspace{-0.5cm}
\hspace{-0.5cm}
\includegraphics[width=0.9\columnwidth]{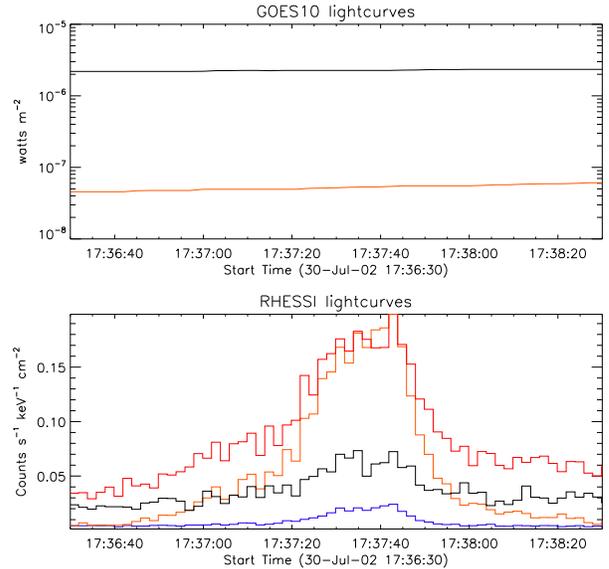} 
\caption{\label{fig_30_jul_2002_over} 
{2002 July 30 flare: 
GOES (3\hspace{0.1cm}s; upper panel) and 
RHESSI (2 second bins) lightcurves  (bottom panel)
in: 3-9\hspace{0.1cm}keV (black), 9-15\hspace{0.1cm}keV (red), 15-30\hspace{0.1cm}keV (orange), 30-100\hspace{0.1cm}keV (blue).}  }
\end{figure}

\begin{figure}[!t]\centering
\includegraphics[width=0.33\columnwidth]{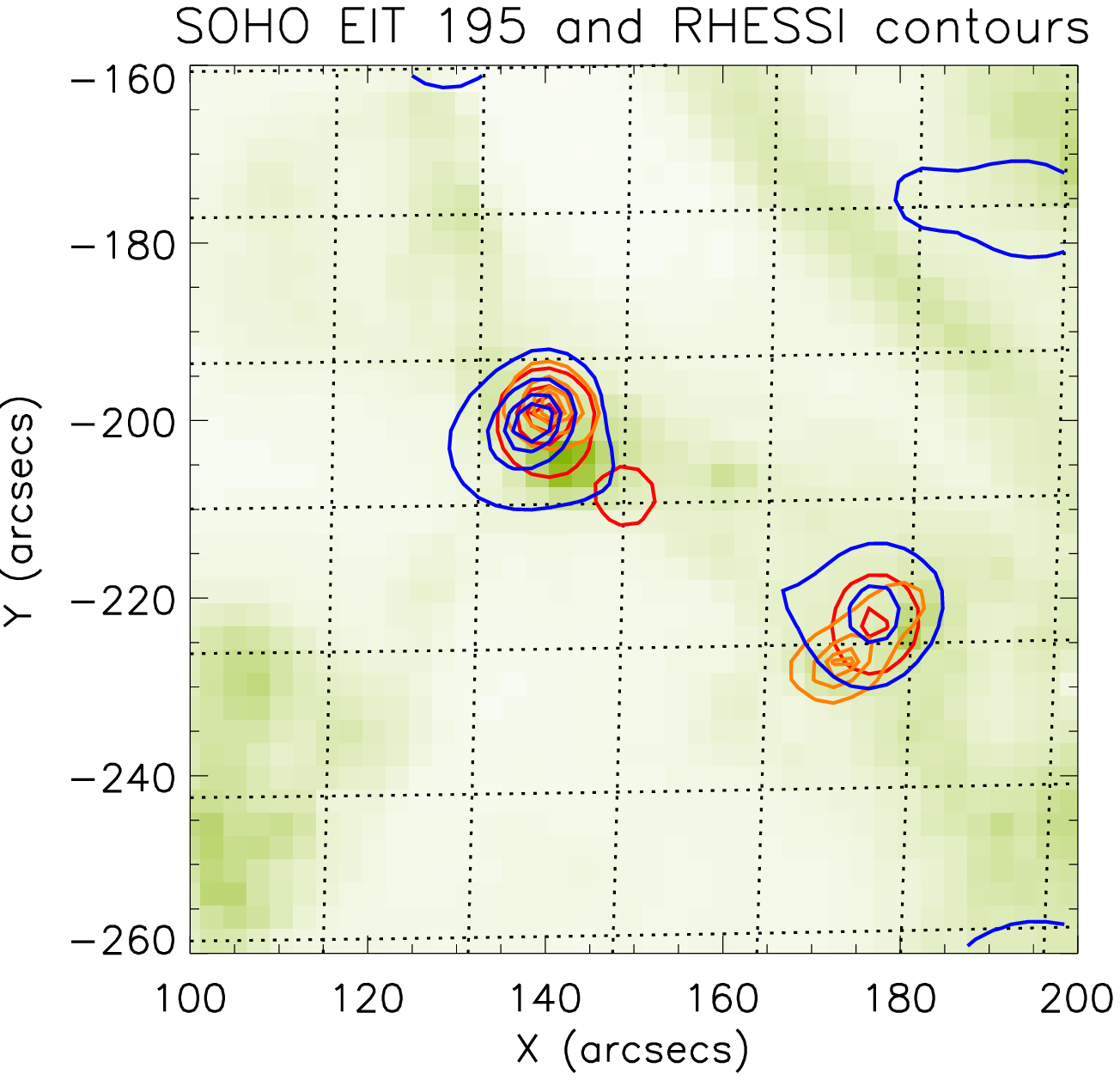} 
\hspace{-0.5cm}
\includegraphics[width=0.31\columnwidth]{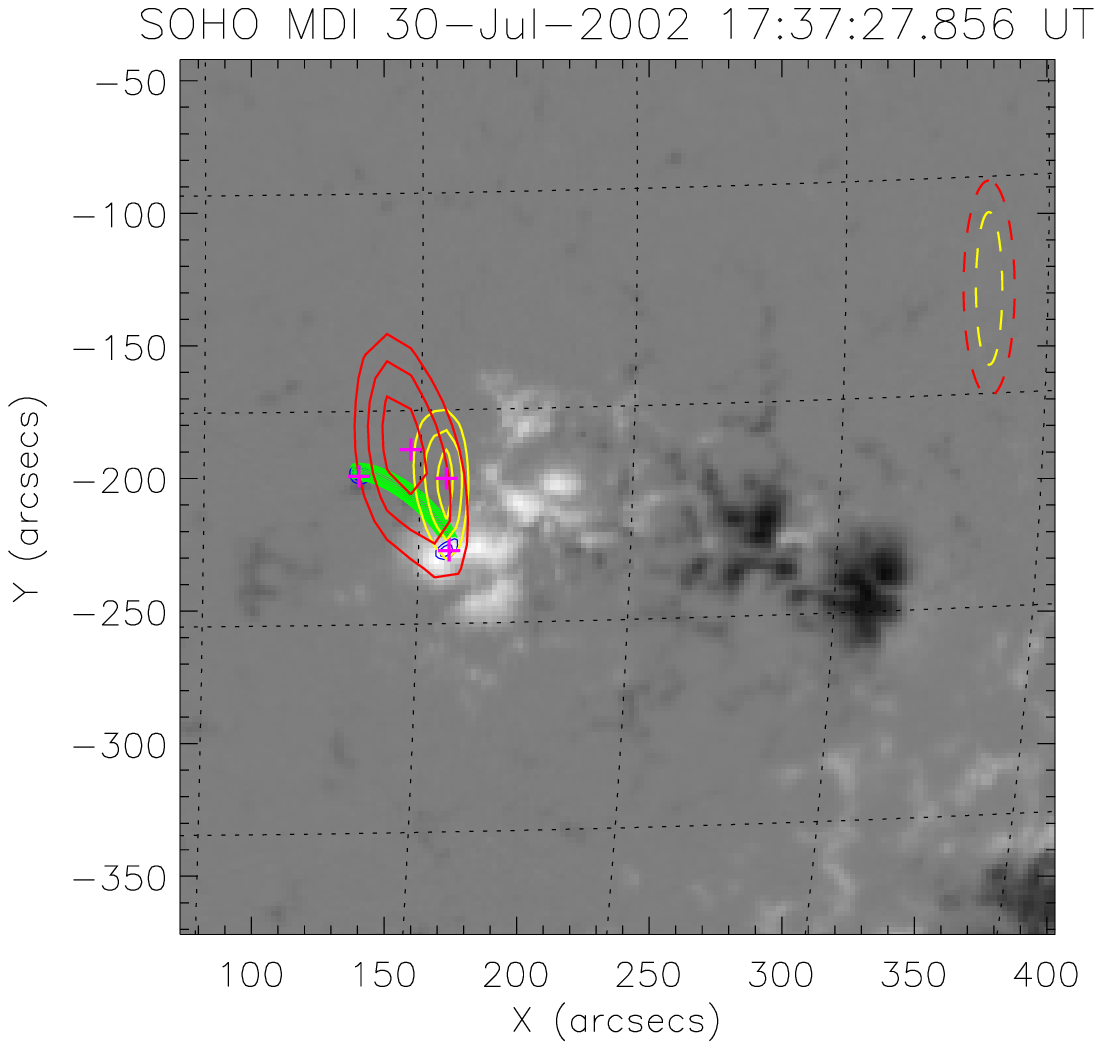}
\includegraphics[width=0.31\columnwidth]{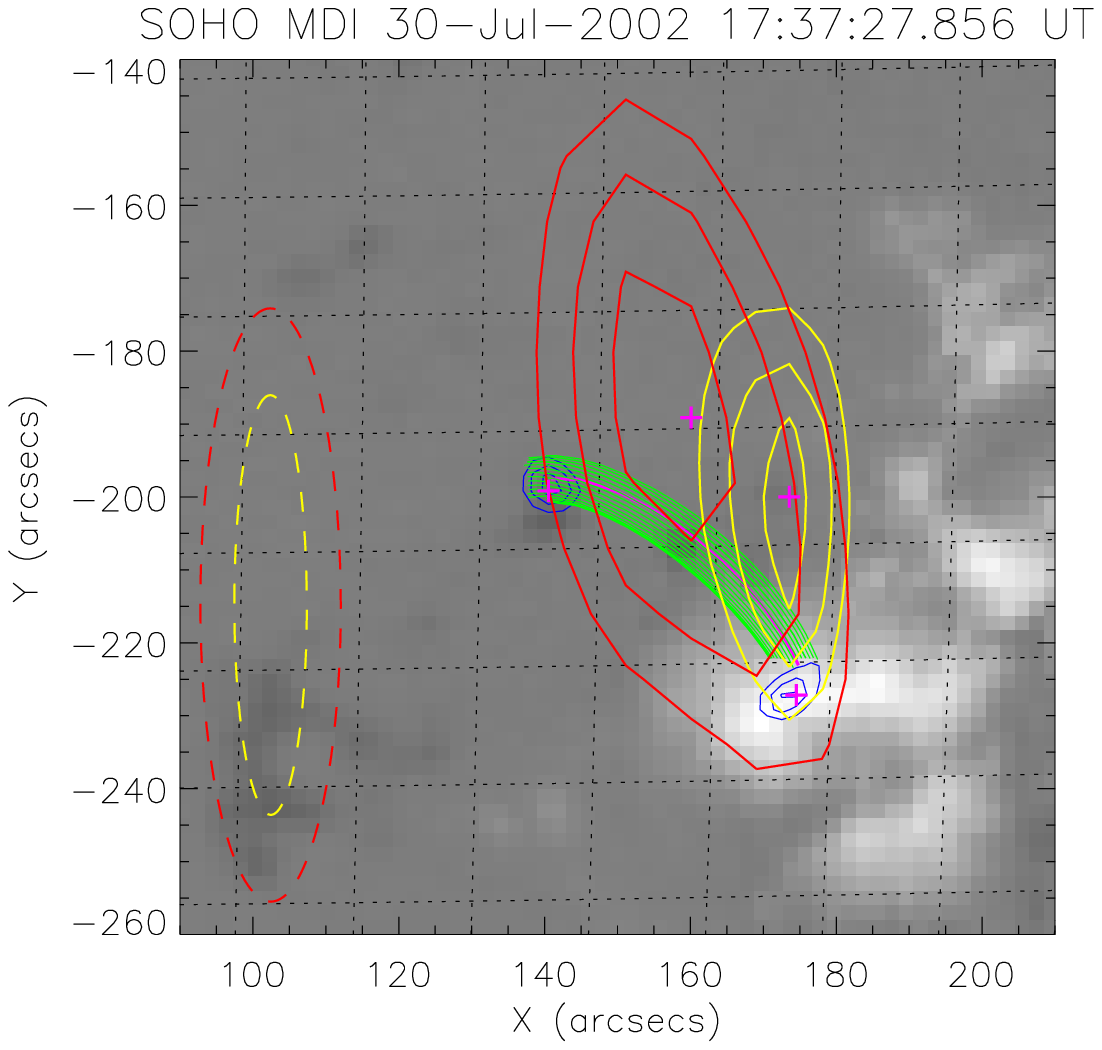}
\caption{\label{fig:Ximage} {Left: Spatial distribution of X-ray emission from 2002 July 30 flare
in various energy ranges contours at 30, 50, 70, 90\% levels: $9-15$\hspace{0.1cm}keV (red) $15-30$\hspace{0.1cm}keV (orange), $30-100$\hspace{0.1cm}keV (blue).
 Background image is SoHO EIT\hspace{0.1cm}195
taken just before the\hspace{0.1cm}flare at $17:36$\hspace{0.1cm}UT. Middle and right: the\hspace{0.1cm}full and close-up view of the\hspace{0.1cm}active region and an extrapolated flux tube (green)
connecting two X-ray footpoints (blue contours),  $2.6-3.2$\hspace{0.1cm}GHz radio image (red contours)
and $4.2-8.2$\hspace{0.1cm}GHz (yellow contours). Magenta plus signs mark the\hspace{0.1cm}spatial peaks of the\hspace{0.1cm}HXR and radio sources. Dashed ellipses display the\hspace{0.1cm}sizes of the\hspace{0.1cm}synthesized beams.}}
\end{figure}

Another vivid example illustrating the unique capacity of microwave observations is a cold, tenuous flares (RHESSI nugget No. 153)---which allowed detection and study of the very acceleration region of fast electrons in the flare. The unusual nature of this event is immediately apparent in Fig.~\ref{fig_30_jul_2002_over}. There is a nice RHESSI hard X-ray burst of about one minute duration, with count rates like those typically seen in M-class flares,  but remarkably there was no counterpart visible at all (less than the GOES C1 level) in soft X-rays. The location and morphology of the event are also interesting. Fig.~\ref{fig:Ximage} shows that it occurred in the following portion of the active region. In this figure the green arc in the right panel shows a magnetic loop identified at the correct location in the extrapolated magnetic model. The three bands shown in the left panel are 9-15 (red), 15-30 (orange), and 30-100 keV (blue). Note the lack of a looptop source at any point in the event---the footpoint structures extend to remarkably low (9-15 keV) energy.  It is otherwise the classical morphology of a coronal flux bundle, with a length inferred to be some $4\times10^9$~cm and volume $6\times10^{26}$~cm$^3$. These geometrical constraints, plus the diagnostic information obtained from the X-ray, radio, and magnetic observations, allow a great deal to be learned about this unusual flare; particularly, about the thermal/nonthermal energy partition with a strongly dominating nonthermal component, which says something remarkable about the acceleration mechanism operating in this event.  The microwave spectra offer further clues.

\begin{figure}[!t]\centering
\includegraphics[width=0.9\columnwidth, bb=77 75 600 435, clip]{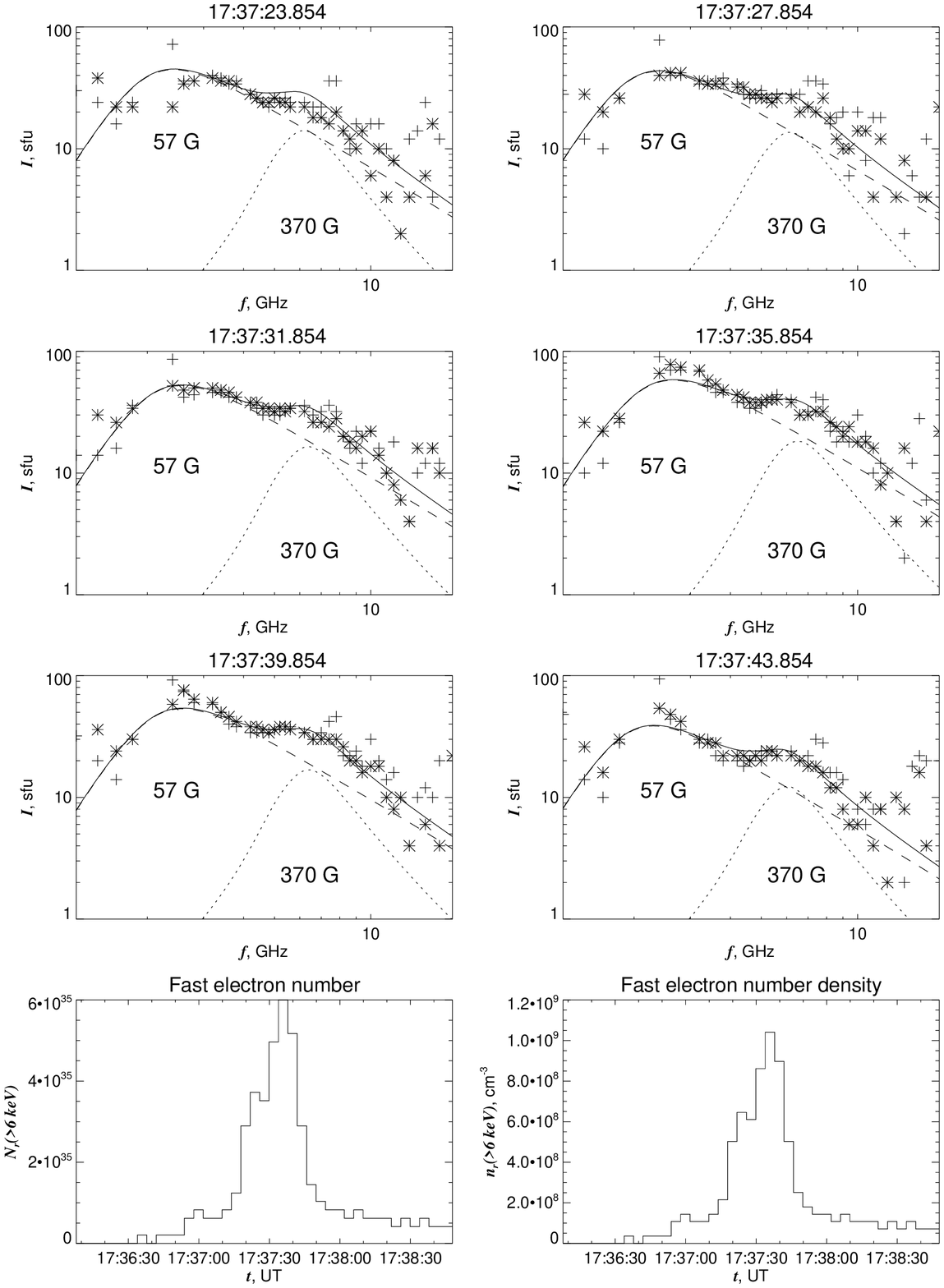}\\
\vspace{-0.25cm}
\caption{\label{fig:fit_tt}
{OVSA radio spectra obtained by two small antennas (pluses and asterisks) and model GS emission from the acceleration region (dashed lines), precipitating electrons (dotted lines), and sum of these components (solid line). Total number and number density of the fast electrons at the radio source as derived from the OVSA radio spectrum.}}
\end{figure}

Fig.~\ref{fig:fit_tt} includes some double-peaked microwave spectra fitted with two components of differing magnetic field strength, and shows the inferred number and number density time profiles of electrons involved in producing the observed radiation (lower panels).  \citet{Fl_etal_2011} find that $6\times10^{35}$ electrons above 6 keV are necessary, and that the density of these electrons approaches 10$^9$~cm$^{-3}$. This density, consistent with the RHESSI-derived acceleration rate for a fast-electron trapping time of 3~s, easily agrees with the low upper limit for the thermal plasma distribution in the loop, inferred from the lack of soft X-rays.  \citet{Fl_etal_2011} conclude that the event was dominated by non-thermal plasma, with the main (low-frequency) component being due to the acceleration site itself, in a region of low magnetic field {($\sim 60$~G)}, {while} the second (high-frequency) component {arises from} the precipitation region of substantially higher magnetic field strength {($\sim 370$~G)}. Although the energetic electrons bombard the chromosphere at a rate comparable to that of an average GOES M-class flare as seen from RHESSI, the flare shows little evidence for significant thermal plasma heating or chromospheric evaporation. A highly asymmetric flaring loop, combined with rather low thermal electron density, have made it possible to detect the \textbf{GS radio emission} directly from the \textbf{acceleration site}. We appear to have here a clear case of \textit{a thermally ``cold" object dominated by nonthermal acceleration}: the electron distribution function in the coronal plasma consists mainly of fast particles, with no evidence for a comparably {strong} thermal signature. The physics of such an object is ill-understood at the present time but is consistent with a \textbf{stochastic acceleration mechanism} with a roughly \textit{energy-independent lifetime} for the fast electrons in the acceleration region. This discovery, therefore, offers stringent new constraints on the acceleration mechanism in flares.  

\section{A "normal", 11 Apr 2002, flare: mild acceleration with a significant heating}

\begin{figure}\centering
\includegraphics[width=0.85\columnwidth]{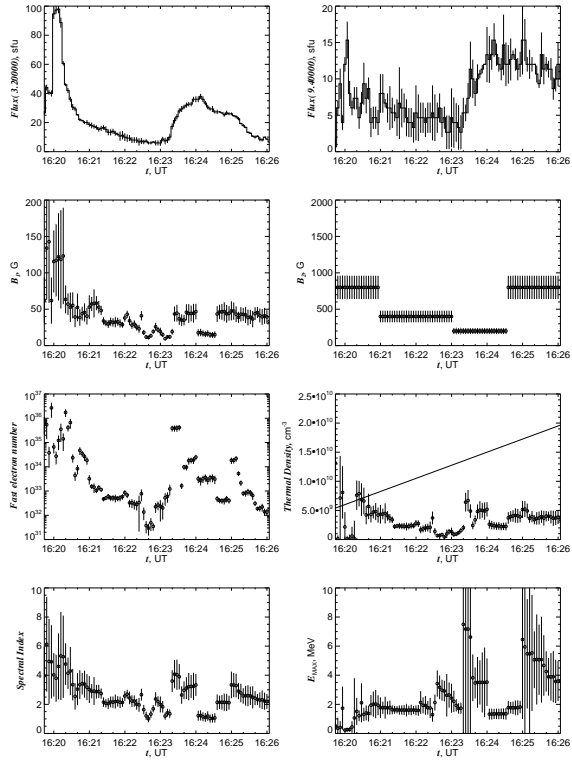}
\caption{\label{fig_OVSA_fit_parms} Radio source parameters as derived from the OVSA spectral fit  for five parameters of the low-frequency coronal source and adopted magnetic field value $B_2$ for the 'precipitating' source as described in the text. A solid curve at the thermal plasma number density shows a number density evolution of the SXR source derived from emission measure from the RHESSI fit. Two top panels show the radio light curves recorded at 3.2~GHz and 9.4~GHz given for the reference purpose. }
\end{figure}

One more example, where the radio contribution from acceleration region dominates the low-frequency part of the microwave spectrum over a limited time is the 11 Apr 2002 flare; Fig.~\ref{fig_OVSA_fit_parms}. The acceleration region contribution was identified with the impulsive peak from 16:20:00 to 16:20:20~UT based on analysis of images and light curves and confirmed by radio spectral fit, the results of which are shown in Fig.~\ref{fig_OVSA_fit_parms}.

\begin{figure}\centering
\includegraphics[width=0.45\columnwidth]{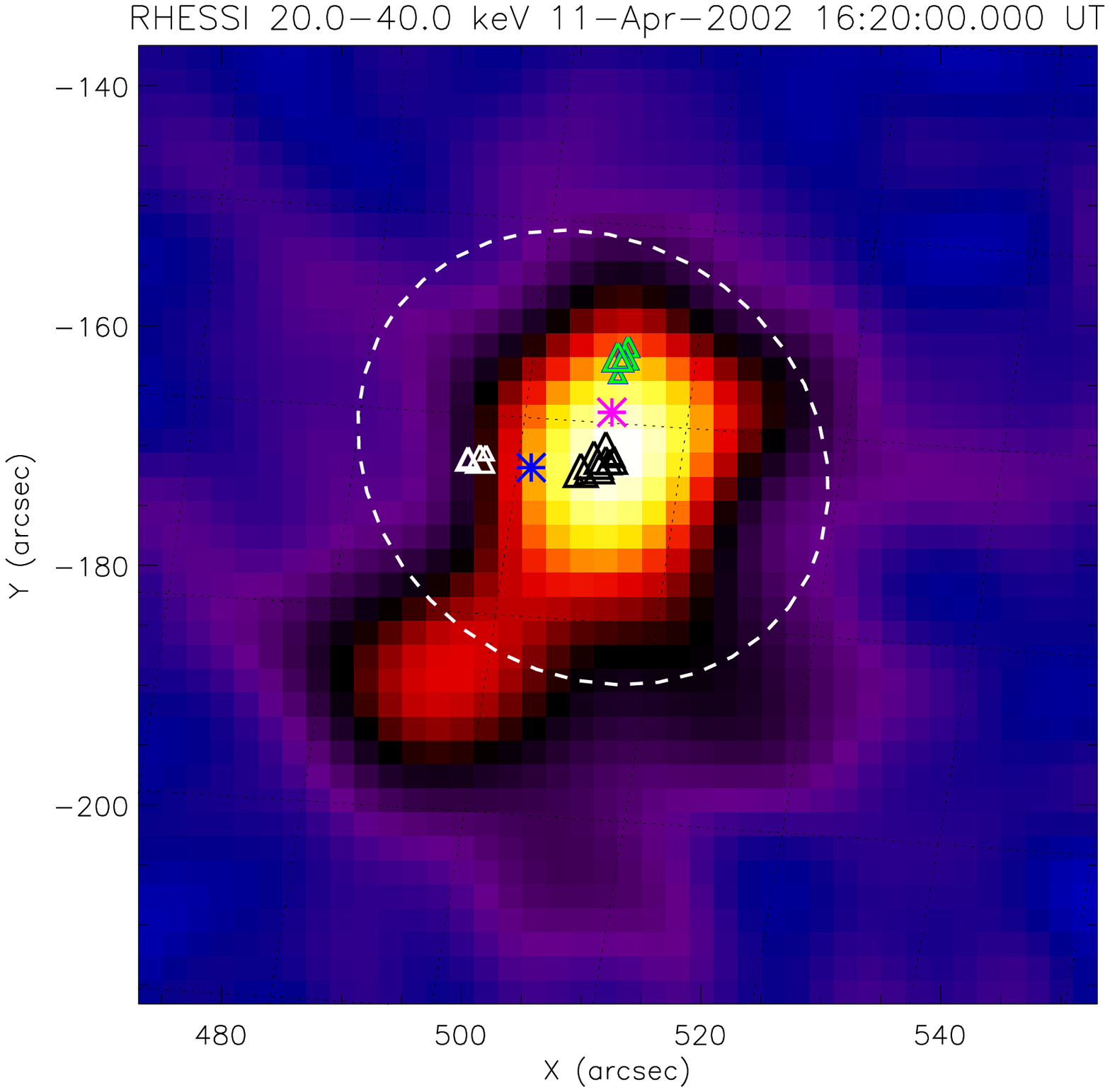}
\includegraphics[width=0.4\columnwidth]{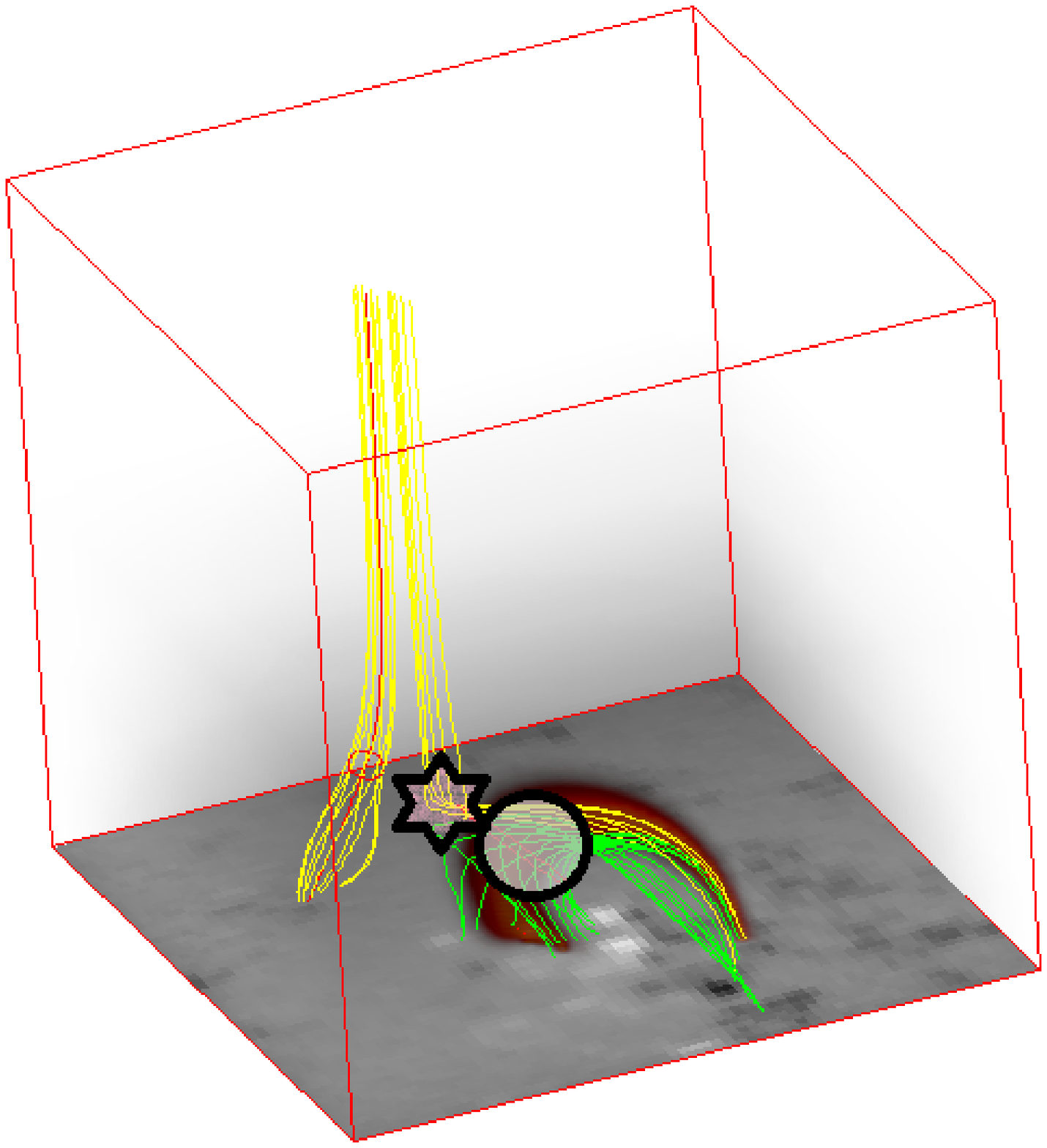}
\caption{\label{fig_3D_set}
 Left: Evolution of the spatial brightness peak of the radio emission at 2.6--4.2~GHz from April 11, 2002 flare.
Background: HXR image at $20-40$~keV at 16:20:00--16:21:00~UT. Symbols are OVSA image centroid positions separated by 8~s time interval (OVSA temporal resolution for imaging):
green triangles are for snapshots from 16:19:55 to 16:20:15~UT, pink asterisk for 16:20:20~UT, black triangles from 16:20:25 to 16:22:00~UT, white triangles from 16:23:40 to 16:24:20~UT, and the blue asterisk is for a late decay phase of 16:25:08~UT; larger triangles correspond to later snapshots within each group. The sequence of the contours clearly indicates that the radio source  is located at the northern part of the HXR image and stays there during the entire impulsive phase, then moves southward to exactly match the HXR centroid position and stays there the entire decay phase of the first peak. The synthesized beam is shown by the dashed white oval.
Right: 3D model of the flaring region based on LFFF extrapolation with $\alpha\approx -5.5\cdot10^{-10}$~cm$^{-1}$ of the photospheric SOHO/MDI magnetogram visualized by two magnetic flux tube (central field lines are red): the first one consists mainly of the closed field lines (green) with a few outer open field lines (yellow), while the other one consists of open field lines only. The locations of the acceleration region and trapping source are shown on top of this structure by the star symbol filled with a semitransparent rose color and the circle filled with a semitransparent light grey color. }
\end{figure}

The {derived evolution of the} physical parameters deserves {some discussion}. The magnetic field {in} the low-frequency source is about 120~G during the impulsive phase of the radio burst, while it drops quickly to 30--50~G at the transition to the decay phase around 16:20:20~UT. Remarkably, this magnetic field change derived from the \textit{spectral fit}, {Figure~\ref{fig_OVSA_fit_parms},} happens at the very same time as the $10''$ shift of the \textit{spatial brightness peak}, {see Figure~\ref{fig_3D_set}, left}. 
This implies that it makes sense to distinguish between these two {spatially distinct} {low-frequency} sources---the very acceleration region (the early source, with $B\sim120$~G, producing the impulsive radio emission) and the  classical looptop radio source (the  later source, with $B\sim40$~G, producing the radio emission {from magnetically trapped electrons} {over} the decay phase) {spatially coinciding with the HXR source}.

\textbf{Acceleration region.} At the {impulsive low-frequency} source, {$\sim$ 16:20:00--16:20:20~UT,} the thermal number density {obtained from the radio fit} is somewhat low, {$n_e \lesssim 2\cdot10^9$~cm$^{-3}$, implying that the {radio} source is located in the corona, not at a chromospheric footpoint}, while the number of nonthermal electrons is consistent with the acceleration rate derived from the HXR data, $(1-3)\cdot10^{34}$~electron/s, {if they reside at the {radio} source for} 2--4~s,  {which requires the strong diffusion transport mode}. The radio derived electron spectral index does not display any significant departure from the HXR derived electron spectral index  {during this time interval}. All these properties are similar to those determined for the acceleration site in the cold, tenuous flare discussed above, {from which we conclude} that we have here another instance of the acceleration region detection in a solar flare.   {The electrons accelerated at this source escape from there in roughly 3~s and then accumulate in another, 'trapping' source, which dominates the radio spectrum and spatial location after 16:20:20~UT. The acceleration, however, continues for a longer time:} we note that the maximum electron energy, $E_{\max}$, displays a monotonic increase from $\sim300$~keV to $\sim2$~MeV over this phase of the burst ($\sim$16:20:00--16:20:50~UT), which is reasonable to interpret as {the} growing of a power-law `tail', i.e., the very process of the electron acceleration. Thus, the  {impulsive low-frequency source dominating the radio emission over roughly 16:20:00--16:20:20~UT},  {which produces fast electrons and  supplies them to the coronal trapping site  until at least 16:20:50~UT,} can confidently be identified with the \textbf{acceleration region} of the flare under study.

\begin{figure}\centering
\centerline{\includegraphics[width=0.4\columnwidth,angle=90]{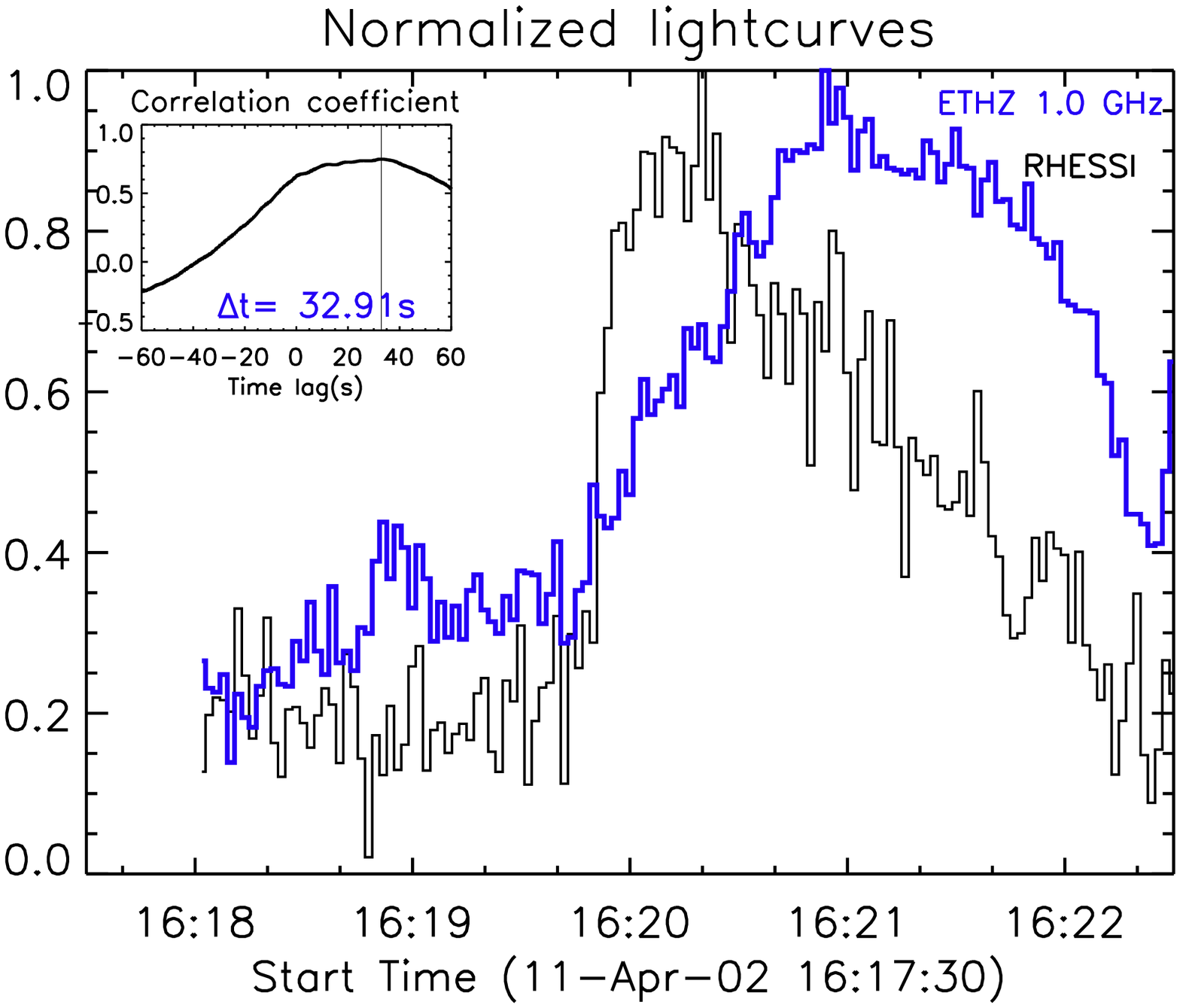}
\includegraphics[width=0.4\columnwidth,angle=90]{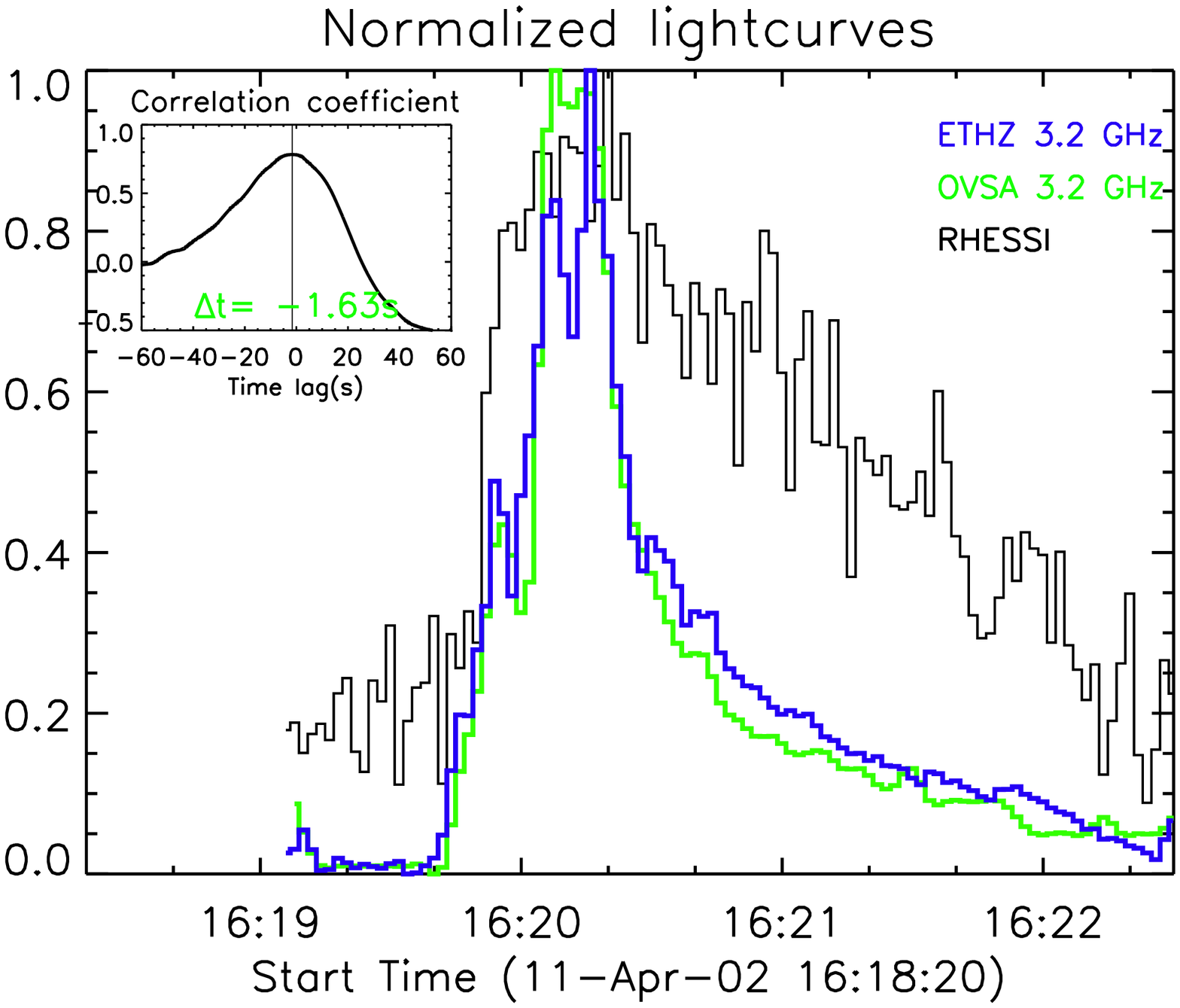}}
\caption{\label{fig:Rtiming_2} Radio to HXR timing: RHESSI 20-40~keV light curve (black) and radio light curves at 1~GHz (left) and 3.2~GHz (right); as observed by Phoenix (blue) and OVSA (green). Insets show the lag-correlation results. for these two cases---non-impulsive light curve on the left and impulsive one on the right.}
\end{figure}

\textbf{Electron accumulation site.}
Transition to the  {gradual} decay  {phase}\footnote{Its main parameters, $B_1\sim 30-50$~G and $n_e \lesssim 5\cdot10^9$~cm$^{-3}$, clearly indicate its coronal location, although spatially distinct from the acceleration region showing a different \textit{coronal} location.} at about 16:20:20~UT  {manifests the stage when the trapping site has accumulated a sufficient number of fast electrons to dominate the radio spectrum. At this time} the derived number of accelerated electrons with $E\gtrsim20$~keV reaches a maximum of $\sim10^{36}$~electrons, which corresponds to the number density of the accelerated electrons of $n_r\sim2\cdot10^8$~cm$^{-3}$ for the adopted source volume. For the RHESSI-derived acceleration rate of $(1-3)\cdot10^{34}$~electron/s, having {a total} electron number {of} $\sim10^{36}$ requires a highly efficient electron trapping with the trapping time longer than 30~s. Indeed, that long trapping time is fully confirmed by the measured delay between the HXR/impulsive radio light curves and 'non-impulsive' radio light curves, some of which are delayed by almost one minute, see, e.g., the 1~GHz light curve in Figure~\ref{fig:Rtiming_2}, left.

The global parameters of the acceleration region determined from the spectral fit and imaging data are $B\sim120$~G and $V\sim6\cdot10^{27}$~cm$^{-3}$, which are, respectively, two times and ten times larger than  in the  cold, tenuous (2002 July 30) flare \citep{Fl_etal_2011}. The residence time of the fast electrons at the acceleration region is $\sim3$~s, which is comparable to that in the cold, tenuous flare, and is much longer than the {free-streaming time} through the acceleration region.  Again \citep{Fl_etal_2011}, this favors {diffusive electron transport due to their scattering by turbulent waves and, thus,} a stochastic 
acceleration mechanism. {The available data is, unfortunately, insufficient to firmly specify the version of stochastic acceleration mechanism (see, e.g., \cite{Petrosian_2012} or \cite{Fl_Topt_2013_CED} for a recent review) operating in the event; however, it does favor those models predicting a roughly energy-independent diffusion time at the source, like in the cold flare event \citep{Fl_etal_2011}}

\section{Early flare phase: heating-dominated energy release}

In some flares the thermal component appears much earlier than
the nonthermal component in X-ray range \citep{Battaglia_etal_2009}.
\citet{Altyntsev_etal_2012}   studied this this 'early flare phase' using microwave observations from various instruments including NoRH, NoRP, SSRT, and RSTN.
Their findings can be summarized as follows.

\begin{figure}
\includegraphics[width=0.45\columnwidth]{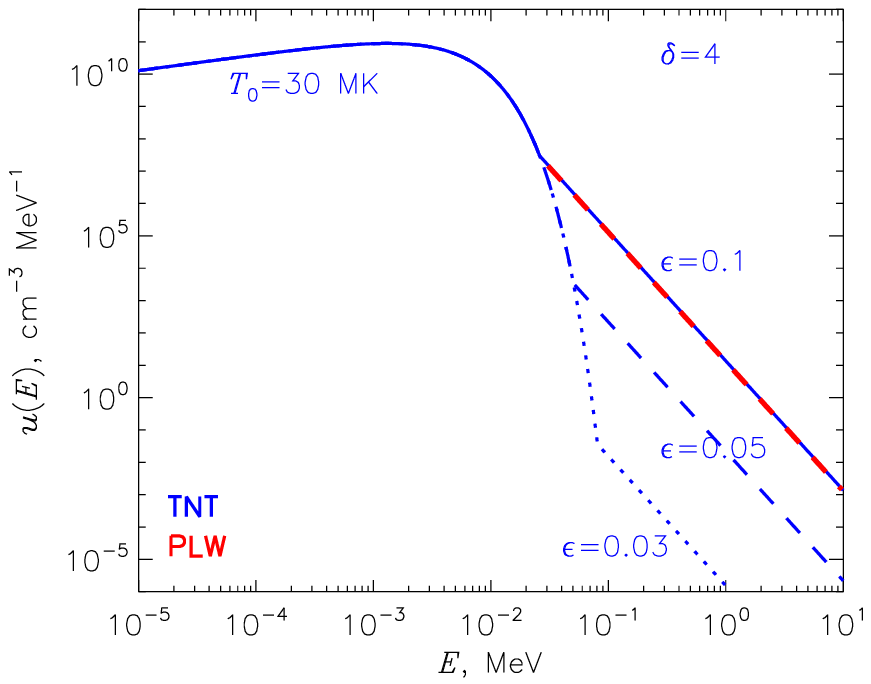}
\includegraphics[width=0.45\columnwidth]{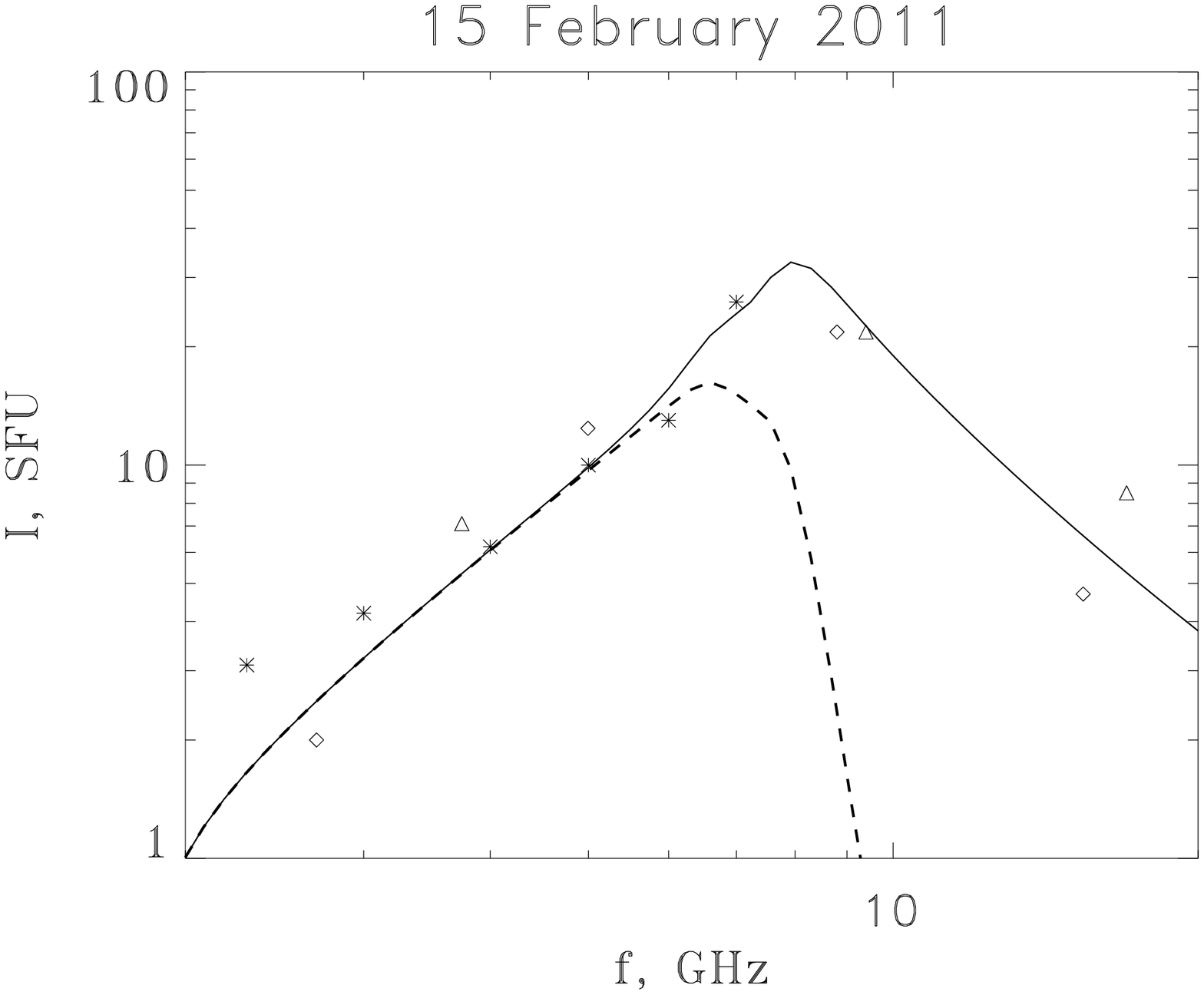}
\caption{Left: Thermal/nonthermal (TNT) electron distribution over kinetic
energy (used to fit the radio spectrum in right panel)  is show for a few different $\varepsilon$ values.
Right: observed microwave spectrum at the preflare phase of the 15 Feb 2011 flare as observed by NoRP, RSTN, and SSRT instruments.
The best TNT fit (solid curve) and purely thermal fit (dashed curve) are superimposed on the observed data.} \label{figTNT}
\end{figure}

First of all, in   all analyzed events there are nonthermal
electrons that generate gyrosynchrotron emission at frequencies
above the spectral peaks. In the case of power-law distribution of
emitting electrons the best fit indices are in the range from 2.5
up to 4, with the high-energy cutoff above 1~MeV. The densities
and energy contents of \textit{nonthermal} components (i.e., above $E_{cr}\sim10-20$~kev, Fig.~\ref{figTNT}) were well below the
\textit{thermal} density and energy of the coronal sources.

Secondly, the thermal GS emission dominates the
\textit{low-frequency} microwave spectra in many cases. This
offers reliable diagnostics of the source area and magnetic
field. The radio estimate of the {coronal} source area is
highly important because it is unbiased by the plasma density
distribution, which is unlike the SXR-derived source area. The
radio data clearly show that the source area grows at
the course of the flare. This fully accounts for the observed
increase of the SXR-derived emission measure, while no density
increase is needed. This means that no essential chromospheric
evaporation occurs in the analyzed cases, so no energy deposition
to the chromosphere in the form of either precipitating electrons
or heat conduction takes place. Since no energy transfer
process is detected,  the early flare phase sources are likely to
represent the energy release and acceleration sites.

Thirdly, even though the thermal plasma contribution to the
microwave spectrum is often essential, no purely thermal stage has
been detected. Indeed,  radio signatures of the nonthermal particles appear
as soon as the plasma heating. Thus, the RHESSI (or Fermi) non-detection of
the nonthermal emission at the early flare phase is accounted by
its relatively low sensitivity, while the microwave observations
turn out to be more sensitive to small numbers of the nonthermal
electrons.


The plasma beta $\beta$ 
is smaller than one, while the nonthermal energy density is much
smaller than the thermal one in all the cases  (i.e., $W_{nth}$
$\ll$ $W_{th}\ll W_{B}$). Although the total energy content of the
accelerated electrons is small, the available nonthermal electrons
are high efficiently accelerated from slightly nonthermal to
relativistic energies. Their spectra are hard, $\delta=2.5-3$ in
most cases, and extended up to a few MeV.  {Stated another
way, the \textit{shape} of accelerated particle spectrum at the
preflare phase is similar to that during flares, even though their
\textit{levels}
(normalizations) are highly different. 
This can happen, for example, if the same, presumably stochastic,
acceleration mechanism capable of accelerating the charged
particles from somehow created \textit{``seed population''} is
involved at both preflare and flare phases. However, these seed
populations must be formed differently in preflares or flares.}

The observed significant
plasma heating suggests that the corresponding flare energy is
already available. However, it is divided highly unevenly between
the plasma heating and nonthermal seed population creation. It works in a way
similar to that in the presence of a DC electric field: at a preflare phase, a
relatively large, but still essentially sub-Dreicer field, will heat
the ambient plasma via the quasi-Joule dissipation (an enhanced, anomalous resistivity is needed to yield a significant plasma heating), while the fraction
of the runaway electrons capable of forming the mentioned seed
population, will remain relatively minor.

 {Let us estimate what DC field $E$ is required to form the seed
populations at the preflare phase. Adopting typical parameters of the preflare source,
$n_{th}\sim 10^{10}$~cm$^{-3}$ and $T\sim30$~MK, the electron Dreicer field is about
$E_{De}\approx3\times10^{-5}$~V/cm. The nonthermal to thermal
electron number density ratio is about $10^{-4}$. To build this
nonthermal component from  the maxwellian tail, electrons  with
$v>v_{cr}$ (where $\exp(-v_{cr}^2/2v_{th}^2)\sim10^{-4}$) must
runaway due to the DC electric field. Given that $v_{cr}^2\sim
(E_{De}/E)v_{th}^2$, we find $E\sim 1.5\times10^{-6}$~V/cm. Over a
typical source size of $\sim10^9$~cm, an electron can gain about
1~keV of energy. This energy is far too small compared with the
observed electron energies of  1~MeV or above. So this assumed DC
field plays no role in forming the nonthermal power-law
distribution responsible for nonthermal GS radiation. However,
this $\sim1$~keV of energy gain can  be sufficient enough to form
a slightly suprathermal seed population, from which the bulk
(presumably stochastic) acceleration produces the observed
nonthermal power-law tails up to relativistic energies.}

Microwave observations, therefore, show that even  the energy release
mechanism in the preflare phase, which is almost thermal, is, nevertheless, accompanied by particle
acceleration.  The nonthermal emission produced by accelerated electrons
with energy of several hundred keV to a few MeV appears as early
as the soft X-ray emission.  The frequency of the spectrum peak is
below 10 GHz for the early flare phase of microwave emission in
all cases, because of a relatively small number of accelerated
electrons at the radio sources. Microwave spectra show that
magnetic field in the coronal sources are a few hundred Gauss at
the early phase. In some cases the number of accelerated electrons
is so low that the gyrosynchrotron emission from \textit{thermal} electrons
dominates the low frequency part of the microwave spectrum.
The microwave observations of
the preflare events are, thus, promising for studying the transitions from the gradual preflare
energy release to the flash flare explosives.



\section{Discussion}

The described findings give rise to a number of fundamentally
important conclusions about the flare heating and acceleration.
(i) The flare energy release is capable of directly heating the
thermal plasma, \textit{without} noticeable \textit{in situ} heating by
fast electron beams or chromospheric evaporation driven by either
electron beams or heat conduction.  (ii) The
fully developed acceleration process of only a minor fraction of the
plasma electrons at the preflare phase implies that the acceleration mechanism involved
is inefficient of accelerating electrons directly from the thermal
pool, but requires a pre-extracted (injected) seed electron
population. (iii) This implies that the electron injection from
the thermal pool and their further acceleration toward higher
energies are driven by physically distinct processes. The first of
them is inefficient or somehow suppressed during the early flare
phase, while the second is already fully operational.
In contrast, at the impulsive phase this injection process is highly efficient up to another extreme, when all or almost all thermal electrons are accelerated.
In
particular, the acceleration by cascading turbulence alone seems
to be insufficient here. Since in the corresponding acceleration
model both injection and acceleration are driven by the same
turbulence intensity, so having these broad range of the thermal-to-nonthermal partitions (from 0 to 100\%) while comparably
efficient acceleration looks at odds to this acceleration model.
(iv) The observed significant plasma heating (in the early flare phases and in 2002 Apr 11 event) suggests that the
corresponding flare energy is already available. However, it is
divided highly unevenly between the plasma heating and nonthermal
population creation.  {We propose, this is due to yet
unspecified energy partition process operating} in a way
 {showing some resemblance to that controlling the balance
between Joule heating, and runaway electrons in a DC electric
field.}
For a larger DC electric field the
fraction of the runaway electrons will grow quickly, resulting in a
powerful nonthermal component needed to produce the impulsive flare phase.
We emphasize, that  {in addition to this energy partitioning
process,}   some sort of stochastic acceleration capable of
producing the observed power-law electron spectra is still needed
during both early and impulsive flare phases.


Properties of the accelerated electron components are somewhat different in the considered 'cold' and 'hot' flares. Firstly, in the 2002 April 11 event the accelerated electron spectrum is noticeably softer ($\delta\approx 5$) than in the cold flares ($\delta\approx 3.5$). Secondly, in the cold, tenuous flare the accelerated electrons are detected at the energies above 6~keV, while in the April 11 event they are only seen above $\sim20$~keV; lower-energy X-ray emission is dominated by the thermal background. Thirdly, the acceleration efficiency is different: in the cold, tenuous flare almost all available electrons were accelerated, while in the April 11 event even the peak instantaneous number density of the fast electrons ($n_r\sim2\cdot10^8$~cm$^{-3}$) does not exceed 10\% of the thermal electron density. Fourthly, in the April 11 event we clearly see a spectral evolution indicative of the growth of a power-law tail ($E_{\max}$ increases with time at the acceleration stage), whereas no spectral evolution was detected in the cold, tenuous flare, which implies a nearly instantaneous growth of the power-law tail. 

Let us discuss from whence all these differences could originate. We have already concluded that the bulk acceleration mechanism is likely to be a stochastic/Fermi process with a relatively long residence time of the electrons controlled by their spatial diffusion on the turbulent magnetic irregularities at the acceleration region. For a diffusive Fermi acceleration process {the shape of the particle energy spectrum} depends primarily on the ratio of two key parameters---the acceleration rate $\tau_a$ {(this is the time needed to establish the nonthermal particle spectrum, not to be interpreted as a duration of the acceleration process)} and the residence/diffusion time of the electrons $\tau_d$ at the acceleration region in such a way that the larger the $\tau_a/\tau_d$ ratio the steeper (softer) the accelerated electron spectrum, (see, e.g., \cite{Hamilton_Petrosian_1992}). The residence times, $\tau_d \sim 3$~s, are comparable in the two events under comparison; the acceleration   {rates} are, however, different. Indeed, the acceleration time $\tau_a$ can be roughly estimated as the time needed for the power-law tail to grow, which is clearly shorter than the residence time, $\tau_a < 3$~s, in the cold flare (recall, no spectral evolution  {was} noted), while longer, $\tau_a > 3$~s, in the April 11 event ($E_{\max}$ increases with time). Thus, for other conditions being equal, the accelerated electron energy spectral index must be larger in the April 11 event in agreement with observations. Note, that the residence time of relativistic electrons in the cold, dense flare is much longer: about 40~s. Interestingly, data on all three flares are consistent with an acceleration process with an energy-independent escape time from the acceleration region.

The acceleration efficiency and energy balance in the flare depend, in addition to the acceleration mechanism itself, on the process of electron extraction from the thermal pool and their injection into the main acceleration process. 
In the cold flare almost all available thermal electrons were injected and accelerated, although their consequent energy losses were insufficient to significantly heat the thermal plasma. In contrast, in the April 11 event, only a relatively minor fraction of the thermal electrons were accelerated, making the collisional heating of the thermal plasma even less efficient than in the cold flare case (given that other relevant physical parameters are similar in these two cases). Thus, the presence of a very hot flaring {SXR} plasma with $T\sim20$~MK (which is present even before the flare impulsive phase) requires another heating mechanism distinct from the collisional plasma heating by accelerated electrons. This conclusion is further supported by the spatial displacement between the thermal SXR source and nonthermal coronal HXR and microwave sources.

Although the available data are insufficient to firmly identify the flare energization process in the presented events, or the mechanism of energy division between the thermal and nonthermal components, we can conclude that this process does show some resemblance to that controlling the balance between Joule heating and runaway electrons in a DC electric field. Indeed, suppose that there is a relatively weak sub-Dreicer electric field directed along the flaring loop magnetic field. This electric field will initiate an electric current, which will lose its energy {through} Joule heating, while the fraction of the runaway electrons available for further stochastic acceleration will be relatively minor. In the case of a stronger electric field, e.g., comparable to the Dreicer field, the fraction of the runaway electrons becomes large, while the Joule heating is reduced so the plasma heating is modest. Even though it is a long way from these speculations to even a qualitative model, the analysis performed favors a flare picture in which electrons are first extracted from the thermal pool by a DC electric field (of yet unspecified origin) and then stochastically accelerated to form a power-law-like energy distribution. Therefore, a stochastic acceleration mechanism naturally containing a DC electric field is called for.

One attractive option naturally providing this combination of the stochastic and DC field acceleration is the case of stochastic acceleration by helical turbulence \citep{Fl_Topt_2013}. Importantly, the turbulence exited on top of a twisted (nonpotential) magnetic field possesses necessarily a nonzero kinetic helicity, which results in a non-zero DC electric field formed by this turbulence. This DC field can efficiently act against the Dreicer field to extract a runaway fraction of electrons from the thermal  pool. This fraction can vary strongly depending on the ratio of this DC field to the Dreicer field, and so can result in any thermal-to-nonthermal energy partition; the runaway electron fraction then  supplies  the stochastic acceleration process. Having the energy-independent escape time is then consistent with nonresonant stochastic acceleration of the electrons by long-wave turbulence \citep{Byk_Fl_2009, Fl_Topt_2013_CED}.

Therefore, we have shown that observations of radio emission directly from the acceleration site provide important constraints on the acceleration mechanism in solar flares.  Despite the great differences between various flares and the preflare phase, a similar acceleration mechanism, although operating in a somewhat different parameter regimes, seems to be called for.  Future radio {spectral imaging} observations that can better separate the acceleration site from the sites of trapping and precipitation are needed to investigate the flare acceleration mechanism(s) in more detail.

{\small This work was supported in part by NSF grants
AGS-0961867, AST-0908344,  AGS-1250374 and NASA grants NNX10AF27G and NNX11AB49G to New Jersey
Institute of Technology and by the RFBR  grants 12-02-00173 and 12-02-00616. This work also benefited from workshop support from the International Space Science Institute (ISSI).}

\bibliographystyle{apj}
\bibliography{xray_refs,fleishman,ms_bib}
\end{document}